\documentclass[12pt]{article}

\usepackage{epsfig}
\usepackage{amsmath}
\usepackage{amsfonts}
\usepackage{graphicx}

\def\appendix#1{
  \addtocounter{section}{1}
  \setcounter{equation}{0}
  \renewcommand{\thesection}{\Alph{section}}
 \section*{Appendix \thesection\protect\indent \parbox[t]{11.715cm} {#1}}
  \addcontentsline{toc}{section}{Appendix \thesection\ \ \ #1}
  }

\newcommand{\br}{\langle}
\newcommand{\kt}{\rangle}
\newcommand{\Zt}{{\Bbb Z}_2}
\newcommand{\de}{\partial}
\newcommand{\rf}[1]{(\ref{#1})}
\newcommand{\non}{\nonumber \\*}

\def\po{\phi_0}

\begin{document}
\begin{titlepage}
\vskip0.5cm
\begin{flushright}
DFTT 5/2001\\
HUB-EP-00/05 \\
ITEP-TH-81/00  
\end{flushright}
\vskip0.5cm
\begin{center}
{\Large\bf Bound states and glueballs in
three-dimensional Ising systems}  
\vskip 0.3cm
\end{center}
\centerline{
M. Caselle$^a$, 
M. Hasenbusch$^b$, 
P. Provero$^{c,a}$ 
and K. Zarembo$^{d,e}$ 
}
\vskip 0.3cm
\centerline{\sl  $^a$ Dipartimento di Fisica Teorica dell'Universit\`a di 
Torino and}
\centerline{\sl Istituto Nazionale di Fisica Nucleare, Sezione di Torino}
\centerline{\sl via P.Giuria 1, I-10125 Torino, Italy}
\vskip .1 cm
\centerline{\sl $^b$ Humboldt Universit\"at zu Berlin, Institut f\"ur Physik}
\centerline{\sl Invalidenstr. 110, D-10115 Berlin, Germany}
\vskip .1 cm
\centerline{\sl $^c$ Dipartimento di Scienze e Tecnologie Avanzate}
\centerline{\sl Universit\`a del Piemonte Orientale, Alessandria, Italy}
\vskip.1cm
\centerline{\sl $^d$ Department of Physics and Astronomy} 
\centerline{\sl University of British Columbia, Vancouver, BC V6T 1Z1, Canada}
\vskip.1cm
\centerline{\sl $^e$ Institute of Theoretical and Experimental Physics,}
\centerline{\sl B. Cheremushkinskaya 25, 117259 Moscow, Russia}
\vskip.2cm
\centerline{\tt E-mail addresses:} 
\centerline{\tt caselle@to.infn.it, hasenbus@physik.hu-berlin.de}
\centerline{\tt provero@to.infn.it, zarembo@physics.ubc.ca}
\vskip.2cm
\begin{abstract}
\par\noindent
We study the spectrum of massive excitations in      
three-dimensional models belonging to 
the Ising universality class. 
By solving the Bethe-Salpeter equation for $3D$ $\phi^{4}$ theory in 
the broken symmetry phase we show that recently found 
non-perturbative states can be interpreted as bound 
states of the fundamental excitation. We show that duality predicts an 
exact correspondence between the spectra of the Ising model in the 
broken symmetry phase and of the $\Zt$ gauge theory in the 
confining phase. The interpretation of the glueball states of the 
gauge theory as bound states of the dual spin system allows us to 
explain the qualitative features of the glueball spectrum, in 
particular, its peculiar angular momentum dependence.
\end{abstract}
\end{titlepage}

\section{Introduction}

The concept of universality is one of the fundamental principles 
in the theory of 
critical phenomena. According to the universality hypothesis,
the long-range behavior of a statistical system 
near a phase transition is governed by fluctuations of 
order parameters
and is essentially determined by symmetries,
no matter how complicated 
the
underlying microscopic dynamics is. 
For systems with $\Zt$ symmetry universality leads to the
description in terms of the one-component $\phi^4$ field theory
(see {\it e.g.} Refs.~\cite{Itzykson:1989sy,Zinn-Justin:1989mi}).
\par
It is known that
an order parameter in three-dimensional 
systems does not fluctuate strongly and 
the mean-field approximation (Landau theory) 
works reasonably well.
The systematic methods that start from the mean-field
approximation and perturbatively
take into account fluctuative corrections give very
accurate predictions for universal quantities such as critical exponents.
However, a detailed numerical study of correlation functions
in the scaling regime of various systems with $\Zt$ symmetry
revealed deviations from the mean-field approximation which are
not too large numerically, but at the same time have no qualitative
explanation within perturbation theory. These deviations
arise in the hierarchy of length scales in the phase with 
broken $\Zt$ 
symmetry.
\par
The only macroscopic scale in the problem, of course, 
is the correlation length $\xi$.
A generic correlator decreases as $e^{-r/\xi}$ at infinity,
but there are some correlation functions that decrease
more rapidly.
Composite
operators in the Ising model whose correlation functions
decay faster than $e^{-r/\xi}$
were constructed in \cite{Caselle:1999tm,Caselle:2000yx}  
and include, as typical examples, operators
of non-zero angular momentum.
What are the length scales associated with these operators?
The answer is simple in the Gaussian approximation,
when 
correlators
factorize: a correlation
function of composite operators that factorizes on 
$n$ irreducible correlators behaves  
as $(e^{-r/\xi})^n$. 
The
characteristic decay rate
of any correlation function 
is therefore an integer multiple of the inverse correlation length.
\par
The fact that fluctuations are not exactly
Gaussian should not change this conclusion.
The ratio of the correlation length to any other length
scale is an integer to any order in perturbation theory! This
can be easily understood from the field theory perspective.
The behavior of a correlation function at infinity is determined by
the
singularities of its Fourier transform in the complex momentum plane.
These singularities are associated with particles described by
the quantum field.
The only singularities that arise in perturbative $\phi^4$ theory are
the single-particle pole associated with 
the
elementary quantum
and the multiparticle cuts. Correlation functions
of operators that couple to the single-particle state decay as
$e^{-\mu r}$, where $\mu$ is the inverse correlation 
length, or the particle mass.
Correlation functions of operators that have zero overlap with
the one-particle state (because of the angular 
momentum conservation, for instance) do not have poles in the
complex momentum plane. Their only singularities are multiparticle
branch cuts that start at
$p^2=-n^2\mu^2$. Consequently, such 
correlators decay as $e^{-n\mu r}$ with some integer $n$.
\par
The numerical simulations clearly show that the above 
picture does not hold
in the phase with broken $\Zt$ symmetry. 
The firmest evidence comes from
the existence of a
spin zero state with mass 
just below the two-particle threshold: $M\sim 1.8\mu$,
strongly supported by numerical results
\cite{Caselle:1999tm,Caselle:2000yx,Lee:2000xn}. 
This and other non-perturbative states
give visible contributions to the quantities which are sensitive to
distances of order of the correlation length. For instance, the spin-spin
correlator in the Ising model near the critical point cannot be fitted
by a simple pole-plus-cut structure suggested by perturbation theory
and definitely contains an extra pole below the cut \cite{Caselle:1999tm}.
\par
The simulations show that  dimensionless ratios like $M/\mu$ become constant 
when approaching the critical temperature. 
Moreover they have 
the same value within statistical errors in two different realizations of 
the universality class: 
the Ising model itself and the lattice regularized $\phi^4$ 
theory.
These two facts allow us
to be confident about the universal character of these
effects. Specifically, we expect the same spectrum to appear in all
realizations of the universality class, as long as one moves away from
the critical point by perturbing with a $\Zt$-symmetric operator
towards the region where the symmetry is spontaneously broken. 
\par
To resolve 
the
aforementioned contradiction, we proposed 
to interpret 
the non-perturbative states in the broken symmetry phase of
$\phi^4$ theory
as bound states of elementary
quanta of the scalar field
\cite{Caselle:2000yx}. We shall develop this idea
further in this paper. 
\par
Let us briefly
recollect some facts about bound states.
 Two-particle bound states show up as 
poles in the Fourier transform of correlation functions
below the branch cut associated with the
two-particle threshold. In three dimensions, the threshold  
singularity is logarithmic, and, though the two-particle cut
comes from one-loop Feynman diagrams and thus is accompanied by 
a coupling constant, the smallness of the coupling is compensated by
the large logarithm. In the vicinity of the threshold, perturbation theory
breaks down. Fortunately, the large logarithms can be resummed.
This resummation substantially changes an analytic structure of
correlation functions near the threshold and
produces a pole that is invisible in perturbation theory.
This pole is associated with a two-particle bound state.
Its position is determined by the Bethe-Salpeter (BS) equation, 
which can be solved order by order in perturbation theory. 
%
\par
In this work we analyze the BS equation for $3D$
$\phi^4$ theory in the broken symmetry phase. The analysis shows that
a bound state of the elementary quanta indeed exists. 
We compute its binding energy at the next-to-leading
order. While the value we obtain at the leading order is in very good
agreement with the Monte Carlo data, the series appear to be 
badly behaved,
so that a reliable
numerical estimate of the subleading corrections to the binding energy
would require the knowledge of a large number of terms in the
series.
\par
The $3D$ $\Zt$ gauge theory is 
related to the Ising model by an exact duality transformation. 
Since
the duality transformation maps the confining phase of the gauge
theory into the broken symmetry phase of the spin
model, all of our considerations apply to the glueball spectrum of the
gauge theory as well. Indeed, we shall prove that duality implies the
exact equality between the spectra of the spin model in the broken
symmetry phase and the gauge theory in the confining phase for all
values of the coupling. 
\par
This correspondence suggests a novel interpretation of the glueballs of
the gauge theory as bound states of the dual spin model, which in turn
allows one to understand the qualitative features of the glueball
spectrum, and in  particular its peculiar angular momentum dependence. 
Bound states with different values of the angular momentum are easily 
seen to appear in the low-temperature expansion of the $3D$ Ising model. 
Simple group-theoretic arguments determine how many 
elementary quanta are necessary to construct a bound state of any given 
angular momentum
and parity.
Assuming that the binding energy is always much 
smaller than the mass of the elementary components, and that this 
description of the spectrum in terms of bound states is conserved 
when one moves 
from the deep low-temperature region to the scaling region, one ends up 
with a prediction for the qualitative features of the glueball 
spectrum in $\Zt$ gauge theory, and in particular for the 
angular momentum dependence of the various masses. This prediction is in very 
good agreement with the numerical data for the glueball spectrum, thus lending 
support to the interpretation of glueballs as bound states of the dual spin 
model. 
\par
The rest of the paper is organized as follows: in Sec.~2 we will 
present the detailed computations for the bound state energy in the 
framework of the BS equation for $3D$ $\phi^{4}$ field theory; in Sec.~3 
we
study the
dependence of the binding energy on the magnetic field. 
Sec.~4 contains a proof of the exact correspondence between the 
spectrum of the $3D$ Ising model 
in the broken symmetry phase
and the  $\Zt$ gauge theory
in the confining phase.
In Sec.~5 we describe in detail the argument, already presented in 
\cite{Caselle:2000yx}, that allows us to explain the angular 
momentum dependence of 
the glueball spectrum of the $\Zt$ gauge theory using the dual 
bound state picture. In Sec.~6 we discuss obtained
results and outline possible future developments.

\section{Bound states in $3D$ $\phi^{4}$:  
BS equation}
We consider $3D$ $\phi^{4}$ theory in the broken symmetry phase. 
The Euclidean action 
is
\begin{equation}
S=\int d^3x
\left[\frac12\de_\mu\phi\ \de_\mu\phi+\frac{g}{4!}(\phi^2-v^2)^2\right].
\end{equation}
The field $\phi$ acquires an expectation value and
the perturbative expansion is performed around the stable vacuum
$\phi=v$. Extracting the fluctuating part,
\begin{equation}
\phi =v+\varphi,
\end{equation}
and setting
\begin{equation}
m^2=\frac{g v^2}{3}
\end{equation}
we write the action as
\begin{equation}
S=\int d^3x\left[\frac12\de_\mu\varphi\ \de_\mu\varphi
+\frac12 m^2\varphi^2+\frac{1}{3!}\,
\sqrt{3 g m^2}\ \varphi^3+\frac{1}{4!}g\ \varphi^4\right].
\end{equation}
The quartic coupling $g$ has a dimension of mass, so the 
dimensionless expansion parameter is the ratio $g/m$.

If the coupling is weak, the bound state lies very close to the threshold.
We shall see that the binding energy is exponentially
small in $g/m$. While the exponent can be easily found 
 \cite{Caselle:2000yx}, 
computing the
pre-exponential
factor requires sufficiently complicated
next-to-leading order calculations.
The calculations
are to certain extent
similar to those of Ref.~\cite{Dashen:1975hd} where the BS 
equation was solved
to the first few orders in perturbation theory for (1+1)-dimensional 
scalar field theory with the
cosine interaction. 

\subsection{Leading order solution}

\begin{figure}[h]
\centering
\includegraphics[width=1.0\textwidth]{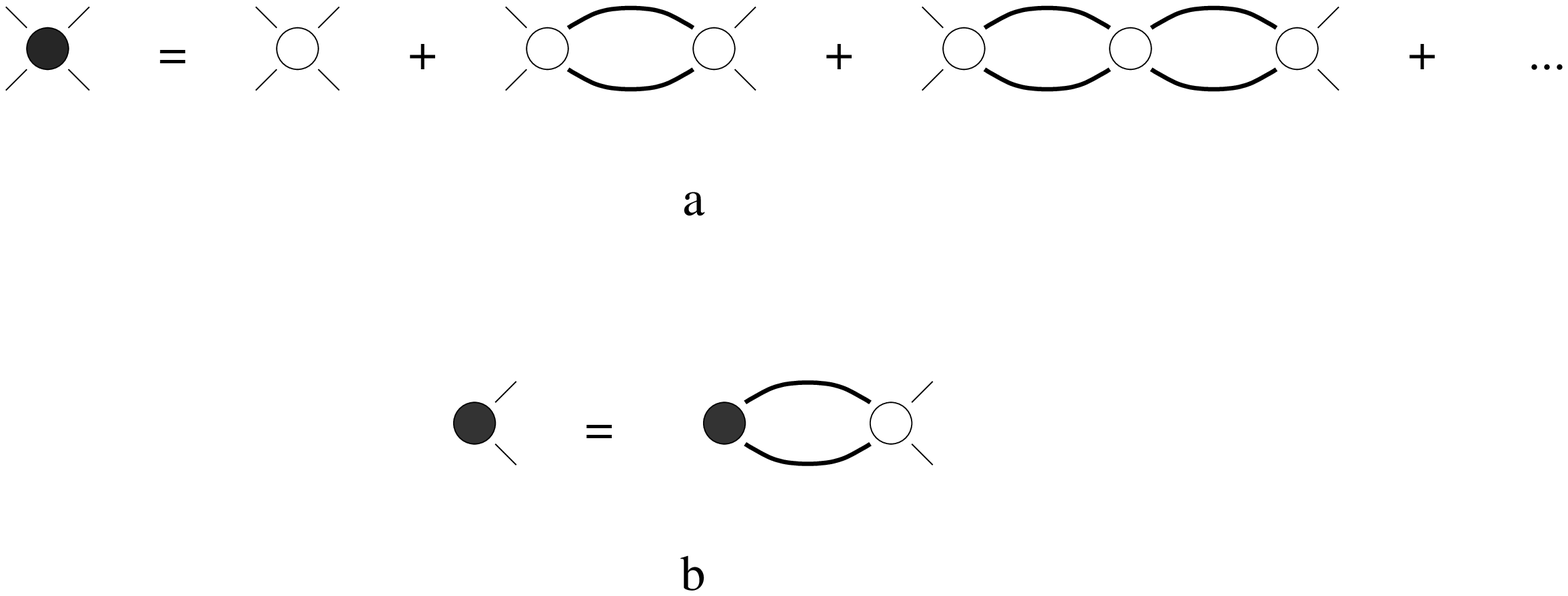}
\caption[x]{\small a) Ladder diagrams. b) BS equation.}
\label{ladder}
\end{figure}

The two-particle bound state manifests itself as a pole in the four-point
function. The pole arises after summation of the ladder
diagrams (Fig.~\ref{ladder}a). Each ladder diagram develops a kinematic
singularity when the momentum flowing through the diagram is near the
two-particle threshold: $p^2\approx-4m^2$, since the momentum integration in
each loop then contains a region where two  
of the intermediate states simultaneously
go on-shell and the propagators in the loop integral
simultaneously have poles. This singularity produces large logarithms that
 compensate for powers of the coupling, so near the threshold  
all ladder diagrams are of the same order.
The vertices 
in Fig.~\ref{ladder}a correspond to the
sum of the two-particle irreducible 
(2PI) diagrams, that is, the diagrams that cannot
be split into disconnected
pieces by cutting two internal lines whose total 
momentum is $p$. 2PI  diagrams are analytic at the two-particle
threshold in the $s$-channel.
\par
The position of the pole is determined by the
BS  equation, graphically represented in Fig.~\ref{ladder}b.
The BS  equation is
 the homogeneous part of the Dyson equation for 
ladder diagrams \cite{Berestetsky:1982aq}:
\begin{equation}\label{bethe}
\Gamma(p_1,p_2)=\int\frac{d^3q}{(2\pi)^3}\,\Gamma(q,p-q)G(q)K(q,p-q;p_1,p_2)
G(p-q),~~~~p=p_1+p_2.
\end{equation}
Here,
 $G$ is the propagator and $K$ is the BS kernel, 
the aforementioned sum of 2PI diagrams. 
The BS equation is a linear eigenvalue problem
for the wave function $\Gamma$ and the bound state mass:
\begin{equation}
M^2\equiv -p^2.
\end{equation}
We take: 
\begin{equation}
M=(2-\Delta)m,
\end{equation}
where $\Delta$ is small.

\begin{figure}[h]
\hspace*{1cm}
\epsfig{file=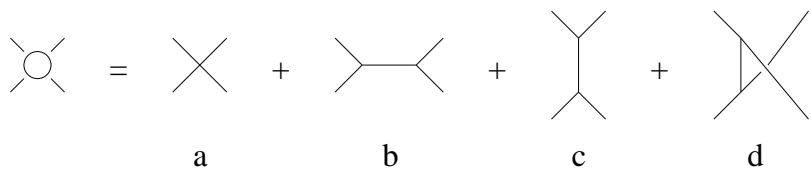,height=2.5cm}  
\caption[x]{\small Tree-level BS kernel.}
\label{kernel}
\end{figure}

To the leading order in $g/m$, the BS 
kernel is given by the sum of three
diagrams in Fig.~\ref{kernel}. The diagram $a$ contributes $-g/2$ to 
the kernel: 
$$K_a=-\frac g2.$$
The diagram $b$ gives 
$$
K_b=\frac{3gm^2}{2} \,\frac{1}{m^2-M^2}=-\frac g2+O(\Delta).
$$
The contribution of the diagrams $c$ and $d$ depends on the momentum transfer
in the $t$ and $u$ channels. However, the momentum transfer is very small when
external momenta and both of the propagators in \rf{bethe} are on shell:
simple kinematic arguments show that 
the
typical momentum
in the $t$ or $u$ channel
propagators is of order $\Delta m$.
To the first approximation, we can just set it to zero:
$$
K_c=\frac{3gm^2}{2}\,\frac{1}{(p_1-q)^2+m^2}\approx \frac{3g}{2}\,, 
$${}
$$
K_d=\frac{3gm^2}{2}\,\frac{1}{(p_2-q)^2+m^2}\approx \frac{3g}{2}\,. 
$$
Collecting all three terms together, we get:
\begin{equation}
K=2g.\label{lk}
\end{equation}
The kernel corresponds to a  local four-point interaction
in this approximation.
The effective quartic coupling appears to be negative.
This happens because
the exchange diagram $c$ dominates over the
 pure four-point vertex $a$ and
the annihilation graph $b$, so the interaction is
attractive. No matter how weak the interaction is, it will bind
elementary quanta into bound states.
\par
Since the binding energy is small, the non-relativistic approximation
should be valid at small $g/m$.
The non-relativistic limit of the BS equation with 
the positive local kernel
is the two-body Schr\"odinger equation with attractive
$\delta$-function
potential. This
quantum-mechanical problem has
a bound state at any value of the coupling with
exponentially small binding energy \cite{Thorn:1979kf}.
An approach based on the non-relativistic approximation
was elaborated in \cite{Caselle:2000yx}. This approach is
simple and physically transparent, but it only allows us to find
the binding energy up to a numerical factor. Here we follow a more
systematic route based on the full BS equation.
\par
The BS 
equation with the local kernel Eq.~(\ref{lk}) corresponds to summing
an infinite series of bubble diagrams, and can be written as
\begin{equation}
1=2g\int \frac{d^3q}{(2\pi)^3}\,\frac{1}{(q^2+m^2)[(p-q)^2+m^2]}. 
\end{equation}
Evaluating the integral, 
\begin{eqnarray}\label{bub}
\int \frac{d^3q}{(2\pi)^3}\frac{1}{(q^2+m^2)[(p-q)^2+m^2]}
&=&\frac{\arcsin\left(\sqrt{\frac{p^2}{4m^2+p^2}}\right)}{4\pi\sqrt{p^2}}
=\frac{\ln\left(\frac{2m+M}{2m-M}\right)}{8\pi M}
\non 
&=&\frac{\ln\left(\frac{4}{\Delta}\right)}{16\pi m}+
O(\Delta\ln\Delta),
\end{eqnarray}
we obtain:
\begin{equation}\label{treebs}
1=2g\frac{\ln\left(\frac{4}{\Delta}\right)}{16\pi m}.
\end{equation}
This equation predicts an exponentially small binding energy:
\begin{equation}\label{be}
\Delta={\rm const}\cdot\exp\left(-\frac{8\pi m}{g}\right), 
\end{equation}
in accord with the result obtained
in the non-relativistic approximation 
\cite{Caselle:2000yx}\footnote{The coupling $\lambda$ used in
\cite{Caselle:2000yx} differs from $g$ by a factor of 24.}.
\par
The normalization of the binding energy  cannot be
deduced from Eq.~\rf{treebs}.
We cannot keep the constant term in
\rf{bub}, because it has the same magnitude as 
logarithmic terms
in the next order of perturbation theory, which are proportional to 
$g^2/m^2\ln\Delta\sim g/m$. There are other sources of corrections 
of order $g/m$, as well. We need to compute all of them to fix the
normalization of the binding energy.

\subsection{Corrections to propagator}

\begin{figure}[h]
\hspace*{1.5cm}
\epsfig{file=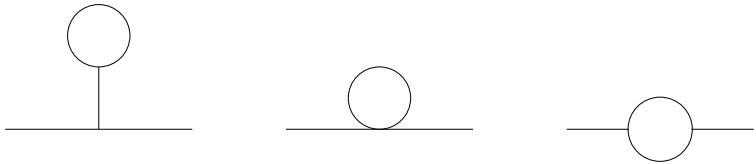,height=10cm,angle=-90}
\caption[x]{\small One-loop corrections to the propagator.}
\label{propagator}
\end{figure}

One-loop corrections to the propagator are given by three diagrams in
Fig.~\ref{propagator}. They shift the pole of the propagator to
\begin{equation}\label{mass}
\mu^2=m^2-\frac{3 g m}{16\pi}\ln 3-g\Lambda,
\end{equation}
where by $\Lambda$ we denote the linearly divergent integral
\begin{equation}
\Lambda\equiv\int\frac{d^3q}{(2\pi)^3}\,\frac{1}{q^2+m^2}.
\end{equation}
Expanding the third diagram in Fig.~\ref{propagator} near the mass shell, we
get:
\begin{equation}
G^{-1}(k)=(p^2+\mu^2)
\left[1+\left(3-\frac{9}{4}\ln 3\right)\frac{g}{24\pi m}\right]
+O\Bigl((p^2+\mu^2)^2\Bigr)
\end{equation}
and
\begin{equation}\label{physprop}
G(k)=\frac{1}{p^2+\mu^2}\,
\left[1-\left(3-\frac{9}{4}\,\ln 3\right)\frac{g}{24 \pi m}\right]
+O(1).
\end{equation}
In evaluating the integral \rf{bub}, 
the corrections to the propagator will result in the replacement 
of the bare mass
$m$ by the renormalized 
(physical)
mass $\mu$ and in the multiplication of the 
whole integrand 
by the wave function renormalization factor.

It is natural to express everything in terms 
of the physical mass gap \rf{mass}. 
In particular, we set:
\begin{equation}
M=(2-\Delta)\mu.
\end{equation}
This definition affects the BS kernel, since
the contribution of the diagram $b$ in Fig.~\ref{kernel}
depends on how exactly we define $M$:
\begin{eqnarray}
K_b&=&\frac{3gm^2}{2}  \frac{1}{m^2-M^2}=
\frac{3g}{2} \frac{m^2}{\mu^2-M^2}\left[1-
\left(\frac{3g m}{16\pi}\,\ln 3+g\Lambda\right)\frac{1}{\mu^2-M^2}\right]
\nonumber\\
&=&-\frac{g}{2}-\frac{g^2}{8\pi m}\ln 3-\frac{2g^2\Lambda}
{3m^2}+O(\Delta)+O(g^3).
\end{eqnarray}
This differs from the result based on our 
previous definition by an amount
\begin{equation}\label{dkb}
\delta K_b=-\frac{g^2\ln 3}{8\pi m}-\frac{2g^2\Lambda}{3m^2}.
\end{equation}

\subsection{One-loop corrections to the kernel}

\begin{figure}[h]
\hspace*{1.5cm}
\epsfig{file=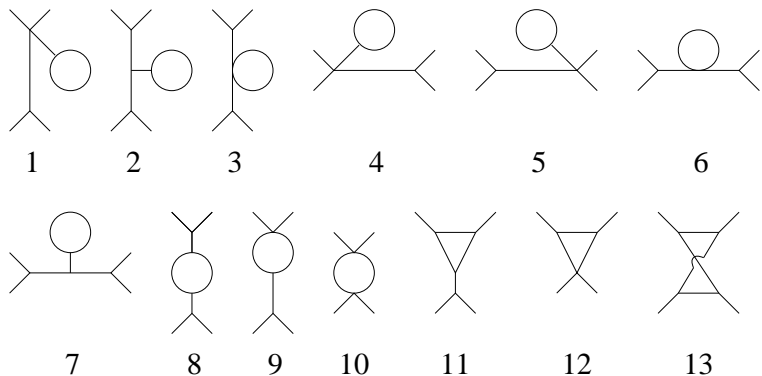,height=10cm,angle=-90}
\caption[x]{\small One-loop two-particle-irreducible diagrams.}
\label{loops}
\end{figure}

The one-loop corrections to the kernel of the BS 
equations are given by
thirteen diagrams in Fig.~\ref{loops}. Since we are
interested only in the logarithmically enhanced terms in  the
BS equation,  all
these diagrams can be evaluated on shell and at zero momentum
transfer in the $t$ and $u$ channels.

The linearly divergent diagrams 1-3 sum up to zero:
\begin{equation}
\delta K_{1-3}=0.
\end{equation}
The diagrams 4-7
cancel the divergence in the diagram $b$, 
the second term in Eq.~\rf{dkb}:
\begin{equation}
\delta K_{4-7}=\frac{2g^2\Lambda}{3m^2}.
\end{equation}
All linear divergences eventually have canceled. It is satisfying
to see that
the net one-loop contribution to the kernel is finite, as it
should be.
 
The remaining diagrams give:
\begin{equation}\label{dk8-13}
\delta K_{8-13}=(9-6+1+12-4+2)\frac{g^2}{16\pi m}=7\,\frac{
g^2}{8\pi m}.
\end{equation}
Collecting together the contributions \rf{dkb}--\rf{dk8-13} we get 
the one-loop correction to the BS kernel:
\begin{equation}
\delta K=(7-\ln 3)\frac{g^2}{8\pi m}.
\end{equation}

\subsection{Momentum dependence}

Finally, we consider the most subtle
corrections that come from  the momentum dependence of
the kernel and of the wave function. Those were
neglected in the leading order calculation because the logarithmic
enhancement of the bubble diagram near threshold comes 
from the region of integration in \rf{bethe}
where all momenta are almost on shell. This allowed us
to forget about the momentum dependence of the wave function
and to 
set
it to one. In fact,
we can always require that the wave function
is equal to the one on shell. This can be regarded as a normalization 
condition.
But to compute non-logarithmic
terms we need to know the complete off-shell wave function.

Once the on-shell condition on the external momenta $p_1,p_2$ is
relaxed, we can no longer neglect the momentum transfer
in the diagrams $c$ and $d$ in Fig.~\ref{kernel}. The BS
 equation then gives:
\begin{eqnarray}
\Gamma(p_1,p_2)&=&
\int\frac{d^3q}{(2\pi)^3}\,\frac{1}{(q^2+m^2)[(p-q)^2+m^2]}
\left[\frac{3gm^2}{(q-p_1)^2+m^2}-g\right]
\non
&=&\frac{g}{8\pi m}\,\ln\left(\frac{1}{\Delta}\right)\left(
\frac{3m^2}{4m^2+p_1^2+p_2^2}-\frac 12\right)
\non
&=&
\frac{3m^2}{4m^2+p_1^2+p_2^2}-\frac 12+O\left(\frac gm\right).
\end{eqnarray}

\subsection{Next-to-leading order BS equation}

With all corrections taken into account, 
the BS equation reads:
\begin{eqnarray}\label{1lbs}
1&=&
\left[1-\left(3-\frac{9}{4}\,\ln 3\right)\frac{g}{24\pi \mu}\right]^2
\non &&\times\int \frac{d^3q}{(2\pi)^3}\,
\left[\frac{3\mu^2}{4\mu^2+p^2+(p-q)^2}-\frac 12\right]
\frac{1}{(q^2+\mu^2)[(p-q)^2+\mu^2]}
\non &&\times
\left[\frac{3g\mu^2}{(q-p_1)^2+m^2}-g+(7-\ln 3)\frac{g^2}{8\pi \mu}\right],
\end{eqnarray}
where external momenta $p_1$ and $p_2$ are on shell: $p_{1,2}^2=-\mu^2$.
We need to retain
terms of order $g/\mu$ and $(g^2/\mu^2)\ln\Delta$ on the right hand side.
For this reason, $O(g/m)$ corrections to the wave function can be neglected,
since they vanish on shell and therefore do not lead to the logarithmic
enhancement. 
The loop integrals encountered in \rf{1lbs} are listed in Appendix and
in \rf{bub}. Using these results, we get after a little of algebra:
\begin{equation}
\log \Delta=-\frac{8\pi \mu}{g}+2\log 2-\frac12 \log 3
+O\left(\frac{g}{\mu}\right)
\label{final}
\end{equation}
that is\footnote{Note that the pre-exponential factor quoted in the published
version of Ref.~\cite{Caselle:2000yx} is not correct.}
\begin{equation}
\Delta=\frac{4}{\sqrt{3}}\exp\left(-\frac{8\pi\mu}{g}\right).
\end{equation}
\subsection{Comparison with numerical results}
To compare Eq.~(\ref{final}) with numerical results one has to know a
value for the dimensionless coupling
$\hat{u}\equiv g/\mu$. The safest thing to do is to relate
$\hat{u}$ to the renormalized coupling $u_R$ defined {\it e.g.} as
in Ref.\cite{Munster:1994qx}, whose critical value is 
precisely known from Monte Carlo
simulations \cite{Caselle:1996wn} 
\begin{equation}
u_R=14.3(1)
\end{equation}
in terms of $u_R$ we have, using results from 
\cite{Caselle:1999tm,Munster:1994qx}
\begin{equation}
\hat{u}=u_R\left[1+u_R\left(\frac{15}{128\pi}+\frac{3\log
3}{32\pi}\right)\right]
\end{equation}
so that 
\begin{equation}
\log\Delta=-\frac{8\pi}{u_R}+\left(\frac{15}{16}+\frac{\log 3}{4}+2
\log 2\right)+O(u_R)
\end{equation}
Therefore the leading-order result is
\begin{equation}
\log\Delta =-\frac{8\pi}{u_R}=-1.76(2)
\end{equation}
corresponding to
\begin{equation}
\Delta=0.172(3) .
\end{equation}
This result is compatible with the assumption of weak binding energy
that underlies the Bethe-Salpeter formalism, and
is in a very good agreement with the numerical result,
\begin{equation}
\Delta_{MC}=0.17(3) .
\end{equation}
However, the parameter of expansion in perturbation theory is 
comparatively large, and the NLO correction is not small:
the $O(1)$ contribution to $\log\Delta$ is $\sim 2.6$, that is
actually greater than the LO contribution. Therefore a reliable
estimate of the subleading corrections to the binding energy seems to
require the
computation of a large number of terms in perturbative series, 
so as to make a resummation possible, which is
typical for application of $3d$ $\phi^4$ theory to critical phenomena.    

\section{Dependence on the magnetic field}

It is possible to study  the dependence of the bound state masses on the
magnetic field. Because the magnetic field couples to the order parameter,
we know how to introduce it in the effective description in terms of $\phi^4$ 
theory. We can 
take the magnetic field  into account by adding a linear term to the action:
\begin{equation}
S=\int d^3x
\left[\frac12\de_\mu\phi\ \de_\mu\phi+\frac{g}{4!}(\phi^2-v^2)^2
-h\phi\right].
\end{equation}
The magnetic field shifts the minimum of the potential, which now
is determined by the equation
\begin{equation}
\frac{g}{6}\,\po(\po^2-v^2)=h.
\end{equation}
Expanding the order parameter around $\po$,
\begin{equation}
\phi =\po+\varphi,
\end{equation}
we get for the action:
\begin{equation}
S=\int d^3x\left[\frac12\de_\mu\varphi\ \de_\mu\varphi
+\frac12 m^2_h\varphi^2+\frac{1}{3!}\,g_3\varphi^3+\frac{1}{4!}\,
g\varphi^4\right]
+S_0,
\end{equation}
where the mass and the trilinear coupling are 
\begin{equation}
m^2_h=\frac{g}{6}\,(3\ \po^2-v^2),
\end{equation}
\begin{equation}
g_3=g\ \po.
\end{equation}
\par
The sum of the diagrams in Fig.~\ref{kernel} gives the
tree-level BS kernel:
\begin{equation}
K_h=\frac{g_3^2}{m^2_h}+\frac{g_3^2}{2}\,\frac{1}{m_h^2-4m^2_h}-\frac{g}{2}
=g\,\frac{7\po^2+v^2}{6\po^2-2v^2}.
\end{equation}
Repeating the same steps as we used to calculate the binding energy
at zero magnetic field,
we get
\begin{equation}
\Delta={\rm const}\,\exp\left(-\frac{16\pi m_h}{K_h}\right),
\end{equation}
where $\Delta$ is the dimensionless binding
energy: $M_h=(2-\Delta)m_h$.

It is useful to introduce the dimensionless variable
\begin{equation}
\xi=\frac{\po}{v},
\end{equation}
which satisfies the equation
\begin{equation}
\xi^3-\xi=\frac{6}{gv^3}\,h=\frac{2}{m^3}\sqrt{\frac{g}{3}}\,h.
\end{equation}
The binding energy is expressed in terms of $\xi$ and $m$, the mass
at  zero magnetic field, as
\begin{equation}
\Delta={\rm const}\,\exp\left(-\frac{8\pi m}{g}\,
\frac{(6\xi^2-2)^{3/2}}{7\xi^2
+1}\right).
\end{equation}
\par
It is easy to see that $\Delta$ is a decreasing function
of $h$. Thus the magnetic field loosens the binding of composite states
and shifts their masses closer to the threshold. 

\section{Duality and the spectrum of Ising systems}
The $3D$ Ising model and  $\Zt$ gauge theory are related by an exact
duality transformation. The broken symmetry phase of the spin model is
mapped into the confining phase of the gauge theory. 
The purpose of this section is to show that, in this phase, 
the duality relationship 
implies an 
exact coincidence of the spectra of the two theories. 
The proof relies on the existence of a non-zero interface tension and
therefore applies exclusively to the broken symmetry phase of the spin
model.
\par
There are several books and reviews which discuss duality in spin systems and 
in particular in Ising-type models  (see {\it e.g.}
Ref.~\cite{Savit:1980ny}). However the duality transformation is
usually treated in the thermodynamic limit only, while to study the
spectrum of the transfer matrix it is necessary to extend the
analysis to finite lattices. This is possible 
as long as one carefully takes 
into account all possible boundary conditions. 
\par
\par
The novel feature of our approach, which greatly simplifies the whole analysis,
is the use of the Transfer Matrix (TM) formalism. 
Let us first recall the definitions of the lattice models we are concerned 
with, and the notion of duality in the thermodynamic limit. 
\begin{itemize}
\item {\bf The spin Ising model}

The Ising model is defined by the action
\begin{equation}
S_{spin} = - \beta_{spin} \sum_{\br n,m\kt} s_n s_m \; , 
\label{Sspin}
\end{equation}
where
the field variable $s_n$ takes the values $-1$ and $+1$; 
$~~n\equiv(n_0,n_1,n_2)$ labels the sites of a simple cubic lattice of size
$L_0$, $L_1$ and $L_2$ in the three directions. 
The 
notation $\br n,m\kt$ in Eq.~(\ref{Sspin})   
indicates that the sum is taken over  
pairs of nearest neighbor sites 
only. 
\item {\bf The gauge Ising model}

The building blocks of the $\Zt$ gauge model are 
the link
variables $g_{n;\mu} \in \{-1,1\}$, which play the role of
gauge fields. 
Denoting by $\mu$  the direction of the link,
the action is
\begin{equation}
S_{gauge} =  - \beta \;\; \sum_{n,\mu<\nu} \;\; g_{n;\mu\nu}
\label{Sgauge}
\end{equation}
where 
$g_{n;\mu\nu}$ are the 
plaquette variables, defined by
\begin{equation}
g_{n;\mu\nu}=g_{n;\mu} \; g_{n + \mu;\nu} \;
g_{n + \nu;\mu} \; g_{n;\nu}~~~.
\label{plaq}
\end{equation}
This action is invariant under local $\Zt$ gauge transformations defined as 
follows: one chooses arbitrarily 
a subset of the sites of the lattice and changes signs 
of all variables defined on the links which end on these sites (if two 
neighboring sites belong to the chosen subset the link that joins them is 
changed twice, that is not changed). It is immediately clear 
that the plaquette values, like any other product of links along a closed 
path, are invariant under this gauge transformation. 
For more details about this model, see {\em e.g.} Ref.~\cite{Drouffe:1983fv}.
\item {\bf Duality} 

There is an exact duality transformation which relates the Ising model and 
the $\Zt$ gauge model. 
This transformation is known as Kramers--Wannier 
duality. It relates the partition functions of the two models evaluated at two
different values of the coupling constants:
\begin{eqnarray}
Z_{gauge}(\beta)~\propto~ Z_{spin}(\tilde\beta)&& \nonumber \\
\tilde\beta&=&-\frac12\log\left[\tanh(\beta)\right]\ , 
\label{dual}
\end{eqnarray}
where $\tilde\beta$ will be denoted as the ``dual coupling''
in what follows.
\par
It is easy to see that low values of $\beta$ are mapped into high values of
$\tilde{\beta}$ and vice versa. Thus the confining region of the gauge
theory is mapped into the broken symmetry phase of the spin model. In
particular the end points of these two phases, the
deconfinement transition and the magnetization transition, are mapped into each
other.
\end{itemize}
An important feature of the dual transformation on  a lattice of finite size is 
that it does not conserve the boundary conditions (BC). In the thermodynamic 
limit this fact becomes irrelevant, but on lattices of finite extent it 
cannot be neglected.
In particular the  $\Zt$ gauge 
model with periodic BC in all directions is mapped by duality into
the Ising spin model with 
{\it fluctuating}
BC, so that
the partition function is given by
the sum of the partition functions with all 
possible choices of periodic ($p$) or antiperiodic ($a$) BC:
\begin{equation}
 Z_{gauge,ppp}(\beta) = \frac12 c^V  (Z_{spin,ppp}(\tilde \beta)
				   +Z_{spin,ppa}(\tilde \beta) +...
				   +Z_{spin,aaa}(\tilde \beta)~)  ,
\label{d1}
\end{equation}
where 
$V\equiv L_0L_1L_2$ is the volume of the lattice, and $c$ is a constant
which can be easily evaluated, but is irrelevant for our purposes. 

This result is discussed in full generality (for a generic lattice geometry and
symmetry group) in~\cite{Gruber:1977}. In the particular 
case of the Ising model on a cubic lattice it can be easily obtained 
by a direct implementation of the duality transformation.
\par
Let us now consider the duality transformation in the framework of the 
transfer matrix approach. The $0$ direction will be our ``time''. 
If we choose periodic BC in the $0$ direction 
without specifying the BC in the $1$ and $2$ directions we obtain:
\begin{equation}
 Z_{spin,pxy}= \mbox{tr} ({\bf T_{xy}})^{L_0} ,
\end{equation}
where ${\bf T}_{xy}$ denotes the transfer matrix of the model in which 
$x,y\in\{a,p\}$ BC are chosen in the 1 and 2 directions respectively.
\par
The antiperiodic BC in the $0$ direction can be obtained by acting with
a spin--flip operator ${\bf P}$, which changes the sign
of all spins in a given time slice. 
Thus we may write
\begin{equation}
 Z_{spin,axy}= \mbox{tr} {\bf P}({\bf T}_{xy})^{L_0} .
\end{equation}
\par
Since the operators ${\bf T}_{xy}$ and ${\bf P}$ commute, they
have a common set of eigenfunctions.
Let us denote the eigenvalues of ${\bf T}_{xy}$ by $\lambda_{xy,i}$ and
those of ${\bf P}$ by
$p_i$. The possible values of $p_i$ are $1$ and $-1$.
States that are symmetric in the magnetization have $p_i = 1$ and
those that are antisymmetric have $p_i = -1$.
\par
Thus we can write
\begin{equation}
Z_{spin,pxy}+ Z_{spin,axy} = \mbox{tr} (1+{\bf P}) ({\bf T}_{xy})^{L_0} = 
2 \sum_{i,sym}
\lambda_{xy,i}^{L_0} ,
\end{equation}
where $\sum_{i,sym}$ means that the sum is restricted
to the states that are symmetric 
in the magnetization
(See  Ref. \cite{Caselle:1994df} for further details 
on this type of construction).
\par
Using this result  we can write the duality relation
in terms of the transfer matrix eigenvalues:
\begin{equation}
\label{important}
 \sum_j \gamma_j^{L_0} = c^{L_0 L_1 L_2} \left[\sum_{i,sym} 
\lambda_{pp,i}^{L_0} 
		       + \sum_{i,sym} \lambda_{pa,i}^{L_0}
		       + \sum_{i,sym} \lambda_{ap,i}^{L_0}
		       + \sum_{i,sym} \lambda_{aa,i}^{L_0}
		       \right] ,
\end{equation}
where $\gamma_j$ are the eigenvalues of the transfer matrix of the
gauge system.
\par
It is instructive to show explicitly that the sums on the two sides of 
Eq.~(\ref{important}) have the same number of terms. 
On the spin side (right hand side 
of Eq.~(\ref{important})) we have  
$2^{L_1L_2-1}$ terms for each sector.
 In fact, the transfer matrix is a $2^{L_1L_2} \times 
2^{L_1L_2}$ matrix, but  only half of the eigenvalues fulfil the symmetry
requirement. Thus taking together all the four sectors we end up with a sum of 
$2^{L_1L_2+1}$ terms.
On the gauge side we have $2L_1L_2$ link variables, but we may fix the gauge.
The maximum number of gauge variables that we may fix, in order to preserve the
periodic BC, is exactly $L_1L_2-1$. This is easy to see: we may fix,
say, all the
links in the $1$ direction except those in the last row, where we may fix
all the links in the $2$ direction, except the last one. Thus we end
up with only
$L_1L_2+1$ degrees of freedom left. Hence the corresponding transfer matrix
has $2^{L_1L_2+1}$ eigenvalues, that is exactly the same number as 
for the spin model.
\par
Since for finite values of $L_1$ and $L_2$ the two sums contain a finite 
number of 
terms
and
since Eq.~(\ref{important}) holds for any integer $L_0$,
each $\gamma_j$ has
to have an exact (up to an overall factor $c^{L_1 L_2}$)
counterpart in the left hand side of the equation. The overall factor 
cancels when one takes the ratios of all the eigenvalues to the lowest 
one to obtain the physical spectrum, and so the latter coincides in 
the two models.
\par
However our goal is to compare the spectra of the gauge theory and the 
spin model {\it both} with periodic BC in all directions, while  
Eq.~(\ref{important}) involves a sum over many different choices of the 
spin model BC. Therefore we must find a relationship between the
$\lambda_{pp,i}$ eigenvalues and
those belonging to the other three sectors. This is easily done by 
noticing that in the low temperature phase of the Ising model 
with antiperiodic 
boundary conditions at least one interface has to be created in the system.
Therefore at leading order (neglecting the cases in which more than 
one interface appear):
\begin{equation}
 Z_{ppa} \approx \exp(- \sigma \; L_0 L_1 ) \; Z_{ppp} ,
\end{equation}
where $\sigma$ is the interface tension.
\par
For the transfer matrix eigenvalues this gives
\begin{equation}
 \lambda_{pp,max}  \approx \exp(\sigma L_1) \;   \lambda_{pa,max} 
\end{equation}
where, with the notation  $\lambda_{pp,max}$ we denote the low lying states of
the spectrum\footnote{
Notice 
that this last argument 
does not hold  in the $2D$ case, where the 
anti-periodic eigenvalues are only suppressed by a constant, independent 
of the system size.}.
\par
So we can conclude that for sufficiently large $L_1$ and $L_2$, 
the low lying spectrum of 
the $Z_2$ gauge theory in the confining phase 
with periodic boundary conditions coincides with the symmetric sector of 
the Ising spin model spectrum in the low temperature phase with periodic 
boundary conditions only. 
(This last argument was already presented in  \cite{Caselle:1996wn}).

\section{The dual bound state picture of glueballs in 
$\Zt$ gauge theory}
In this section we show that the interpretation of the spectrum of 
three-dimensional models in the universality class of the $3D$ Ising 
model as bound states of the elementary quanta provides an analytical 
tool to explain the qualitative features of the glueball spectrum of 
$\Zt$ gauge theory, and especially its angular momentum dependence.
\par
Consider first the spin Ising model in the low-temperature region. Here the
existence of bound states can be immediately inferred from the diagrammatics 
of the low-temperature expansion. This was shown in \cite{Luscher:1988ek} for
the $4D$ case, where however all bound states are expected to disappear in 
the continuum limit due to triviality. 
\par
The mechanism is very simple, and is best explained in the transfer matrix 
formalism. As discussed in \cite{Fisher:1971, Camp:1973}, to the leading order 
in the low-temperature expansion of the transfer 
matrix, 
all time slices are forced to have the same configuration
of spins, and the
the eigenvectors are given by all the possible spin
configurations in a single time slice. The eigenvalues of the Hamiltonian 
(that is minus the logarithm of the transfer matrix) 
are given by the energy of 
each configuration. 
\par
The ground state then corresponds to
the configuration in which all the spins on any time 
slice are parallel; the first excited state corresponds to one flipped spin, 
which requires four bonds to be frustrated. 
To proceed, one has to flip two spins: if these are chosen in non-neighboring 
sites, the energy is just twice the one of the first excitation, since eight
bonds will be frustrated. However one can flip two neighboring spins at the 
cost of frustrating six bonds only: the corresponding state is a bound state 
of the fundamental excitation, with mass just below the two-particle threshold.
\par
One can also construct states of given angular momentum and parity by 
choosing linear combinations of time slice configurations with nontrivial 
transformation properties under spatial rotations and parity reflections.
This can be done systematically by using standard group theory results, 
namely the theory of representations of the dihedral group $D^4$, which is 
the relevant group on a square lattice. This analysis is performed
in Ref.~\cite{Agostini:1997xy}, to which we refer for details.
\par
\begin{figure}
\centering
\includegraphics[width=0.8\textwidth]{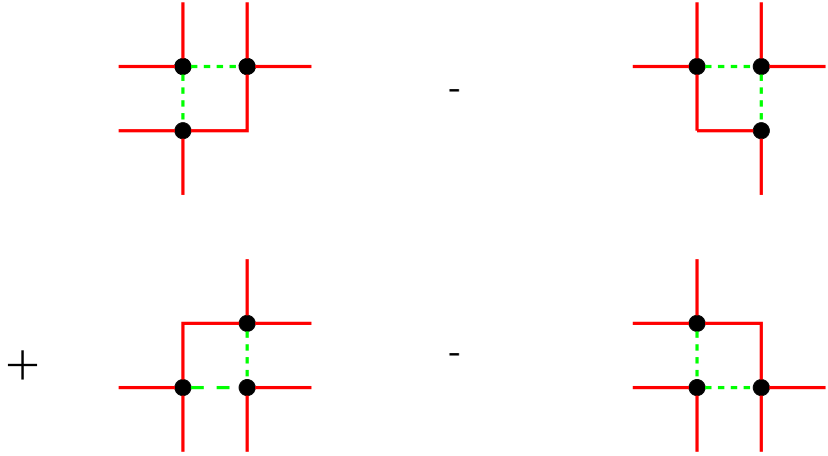}
\caption[x]{\small Three elementary quanta are required to construct a state
with quantum numbers $2^-$. Dots represent flipped spins, solid lines are 
frustrated links, dashed lines satisfied links.}
\label{Fig:nc3}
\end{figure}
\begin{figure}
\centering
\includegraphics[width=1.0\textwidth]{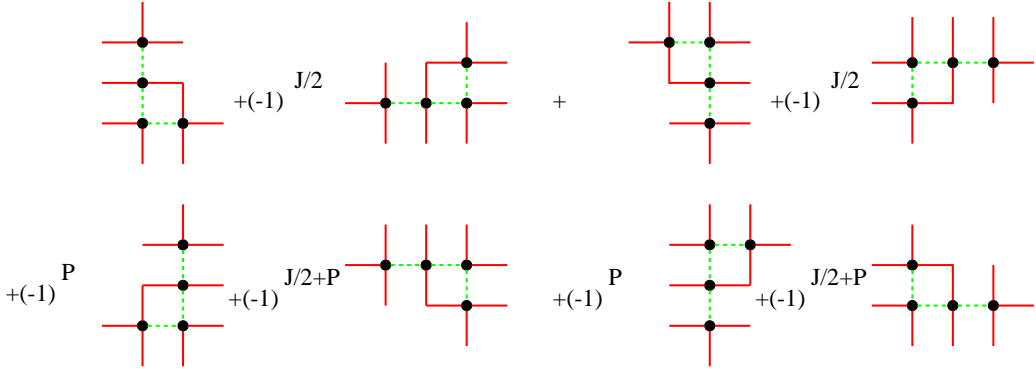}
\caption[x]{\small Bound states of $n_c=4$ elementary quanta.}
\label{Fig:nc4}
\end{figure}

The important point is that any given angular momentum requires a 
certain minimum number $n_c$ of spin flips. 
This means that a bound state of such angular momentum will be composed of at
least $n_c$ elementary quanta. In Tab.~\ref{Table:strong} we have reported 
the values of $n_c$ for the various values of the angular momentum. 
For example,
Fig.~\ref{Fig:nc3} shows the simplest lattice operator corresponding
to angular momentum $J=2$ and parity $P=-1$, which requires $n_c=3$
elementary quanta, while
Fig.~\ref{Fig:nc4} shows how states with quantum numbers $J=0,2$ and
$P=+,-$  can be
constructed with four elementary quanta.  
If we assume that the binding energy is always much smaller that the 
common mass of 
the elementary constituents of the bound states, we end up with a prediction 
for the angular momentum dependence of the glueball spectrum of the 
$\Zt$ gauge theory in the strong coupling regime.
\par
It is not at all guaranteed, of course, that the spectrum will retain these 
same features when going from the strong coupling regime to the scaling 
region: however the data from Monte Carlo simulations of the gauge theory 
show that this is actually the case. The third column of 
Tab.~\ref{Table:strong} contains the Monte Carlo results
for the glueball masses (from Ref.~\cite{Agostini:1997xy}), normalized
to the mass of the lowest glueball. 
This shows
that our dual bound state picture indeed explains the peculiar
qualitative features of the glueball spectrum. 
\begin{table}[h]
\label{Table:strong}
\caption{\sl The bound states that can be constructed with $n_c$
elementary quanta are classified in terms of their angular momentum
and parity. The last column shows the mass of the corresponding
glueball state in $\Zt$ gauge theory.}
  \begin{center}
\begin{tabular}{|c|l|c|}
\hline
$n_c$ &  $J^P$  & $m(J^P)/m(0^+)$\\
\hline
1   &  $0^+$         & 1\\
\hline
2   &  $0^{+*}$      & 1.88(2)\\
\hline
3    &  $0^{+**}$     & 2.59(4)\\
\cline{2-3}
     &  $2^{\pm}$     & 2.59(4)\\
\hline
4    &  $2^{\pm *}$   & 3.23(7)\\
\cline{2-3}
     &  $0^-$         & 3.24(16)\\
\hline
5    &  $0^{- *}$     & 4.48(20)\\
\cline{2-3}
     &  $1^{\pm}$     & 4.12(17)\\
\hline
\end{tabular}
  \end{center}
\end{table}
\par
In particular, some
characteristic approximate degeneracies in the spectrum find a natural
explanation in this picture: for example the states $0^{+,**}$ and
$2^{\pm}$ are
nearly degenerate because they are both bound states of $n_c=3$
elementary constituents. Bound states of $n_c=4$ constituents give 
rise to the $0^-$ and $2^{\pm,*}$ states, again explaining their near
degeneracy, and the same applies to $0^{-,*}$ and $1^{\pm}$ when
interpreted as $n_c=5$ bound states\footnote{The $2^{+}$ state could
be realized with $n_c=2$, but a general theorem (see
Ref.~\cite{Agostini:1997xy}) forces all states with $J\ne 0$ to be
degenerate in parity.}. We do not expect these degeneracies to be
exact, since there is no reason to expect the binding energies to be
exactly the same for bound states of different angular momentum.
\par
Once the minimum number $n_c$ of constituents for a given value of the 
angular momentum has been reached, it is easy to see that states with the 
same angular momentum can be constructed out of any number $n>n_c$ of 
constituents. This suggests that the approximate degeneracies just discussed 
appears between pairs of states only because, in general, only two states can 
be detected with numerical methods in each channel. It is very likely that 
the degeneracies actually group together larger and larger sets of states as 
the number of constituents is increased. For example, we might conjecture 
that the states $0^{-,*}$ and $1^{\pm}$ ($n_c=5$) are nearly degenerate with 
the (as yet undetected) states 
$0^{+,****}$, $2^{\pm,**}$. This degeneracy pattern is a prediction of the 
dual bound state picture of glueballs.
\par
\section{Summary and further developments}
The broken symmetry phase of $3D$ models with $\Zt$ symmetry
shows a rich spectrum of massive excitations, contrary to naive
expectations based on the perturbative field-theoretical
treatment. The non-perturbative states in the spectrum can be interpreted
as bound composites of the elementary particle excitation. When
applied to $3D$  $\Zt$ gauge theory, this approach gives 
a nice explanation of the angular momentum dependence of the glueball
spectrum. In particular, an observed rather peculiar degeneracy pattern
naturally arises in the bound-state interpretation. 
\par
It would be interesting to consider the contribution of the bound states to 
universal amplitude ratios. The obvious way to do that is to use the ladder
approximation for correlation functions, which incorporates all
diagrams that blow up near the two-particle threshold. It would be also
interesting to understand if there are bound states in the systems
with $N>1$ order parameters. The physics is completely different in this
case because the dominating forces are long-range due to the Goldstone
bosons of the broken $O(N)$ symmetry.

\subsection*{Acknowledgements}
K.Z. would like to thank P.~Simon for discussions.
The work of K.Z. was supported by
NSERC of Canada, by Pacific Institute for the Mathematical Sciences
and in part by RFBR grant
98-01-00327 and RFBR grant
00-15-96557 for the promotion of scientific schools.    

\setcounter{section}{0}

\appendix{}
\begin{equation}
\int\frac{d^3q}{(2\pi)^3}\,
\frac{1}{[4\mu^2+q^2+(p-q)^2](q^2+m^2)[(p-q)^2+m^2]}
=\frac{1}{32\pi \mu^3}\,\ln\left(\frac{4}{9\Delta}\right).
\end{equation}
\begin{equation}
\int\frac{d^3q}{(2\pi)^3}\,
\frac{1}{(q^2+m^2)[(p-q)^2+m^2][(q-p_1)^2+m^2]}
=\frac{1}{16\pi \mu^3}\,\ln\left(\frac{4}{9\Delta}\right).
\end{equation}
\begin{eqnarray}
\int\frac{d^3q}{(2\pi)^3}\,
&&\frac{1}{[4\mu^2+q^2+(p-q)^2](q^2+m^2)[(p-q)^2+m^2][(q-p_1)^2+m^2]}
\non
&&=\frac{1}{32\pi \mu^5}\,\left[\ln\left(\frac{4}{9\Delta}\right)
-\frac 23\right].
\end{eqnarray} 
All integrals are calculated on shell: $p_{1,2}^2=-\mu^2$, and near
the two-particle threshold:
$p^2\equiv(p_1+p_2)^2=-M^2=(2-\Delta)^2\mu^2$.

\end{document}